\pdfoutput=1

\documentclass[reqno]{amsart}
\RequirePackage[l2tabu, orthodox]{nag} 

\usepackage{hyperref}

\usepackage{courier}
\usepackage{color}
\usepackage{amsmath,amsthm,amssymb}
\usepackage{cases} 
\usepackage[cal=boondox,scr=boondoxo]{mathalfa}

\usepackage{graphicx}
\usepackage[figurename=Fig.,labelsep=period]{caption}

\usepackage{textcomp} 
\usepackage{sistyle}
\usepackage{physics}
\usepackage{mathtools}
\mathtoolsset{showonlyrefs=false}

\theoremstyle{remark}

\newtheorem*{remark*}{Remark}

\usepackage[]{natbib}
\bibpunct[,]{(}{)}{,}{a}{}{,}

\newcommand{\citetapos}[1]{\citeauthor{#1}'s \citeyearpar{#1}}

\begin{document}

\title[Broken symmetry of L{\'e}vy-stable fluctuations]{Broken symmetry of recruitment fluctuations in marine fishes: L{\'e}vy-stable laws and beyond}
\author[H.-S. Niwa]{H.-S. Niwa}
\date{}


\begin{abstract}
Recruitment is calculated by summing random offspring-numbers entering the population, where the number of summands (i.e. spawning population size) is also a random process. A priori, it is not clear that individual reproductive variability would have a significant impact on aggregate measures for monitoring populations. Usually these variations are averaged out in a large population, and the aggregate output is merely influenced by population-wide environmental disturbances such as climate and fisheries. However, such arguments break down if the distribution of the individual offspring numbers is heavy-tailed. In a world with power-law offspring-number distribution with exponent $1<\alpha<2$, the recruitment distribution has a putative power-law regime in the tail with the same $\alpha$. The question is to what extent individual reproductive variability can have a noticeable impact on the recruitment under environmentally driven population fluctuations. This question is answered by considering the L{\'e}vy-stable fluctuations as embedded in a randomly varying environment. I report fluctuation scaling and asymmetric fluctuations in recruitment of commercially exploited fish stocks throughout the North Atlantic. The linear scaling of recruitment standard deviation with recruitment level implies that the individual reproductive variability is dominated by population fluctuations. The totally asymmetric (skewed to the right) character is a sign of idiosyncratic variation in reproductive success.
\end{abstract}

\maketitle

\section{Introduction}
Marine populations of mass-spawning species are characterized by intermittent, large recruitment events
\citep{Hjort1914}.
Very large interfamilial, or idiosyncratic, variation in reproductive success has been documented in marine species
\citep{Hedgecock-Pudovkin2011}.
Population models with skewed offspring-number distributions have recently been proposed as appropriate models to investigate gene genealogies for abundant marine species with type-III (exponential) survivorship curves
\citep{EW06,Sargsyan-Wakeley08}.
Patterns of genetic variation have been studied to search for the imprint of multiple mergers of ancestral lineages
\citep{SBB13,EBBF2015,Arnason-Halldorsdottir2015}.
The `imprint' is a sign of the domination by a single or a few families in the population.
\citet{Niwa-etal2016} showed that the multiple-merger coalescent model of the $\mathrm{Beta}(2-\alpha,\alpha)$ type provides a better fit of Japanese sardine genetic variation than \citetapos{Kingman82-the-coalescent} classical coalescent model.
The $\mathrm{Beta}(2-\alpha,\alpha)$ coalescent arises from a population model with power-law offspring-number distribution with exponent $1<\alpha<2$
\citep{Schweinsberg2003}.
While the recruiting process has exponential decay in survival probability, the exponential amplification of the number of successfully recruiting offspring (littermates or siblings) in a family compensates the exponentially small probability of their surviving to reproductive maturity.
The combination of these two exponentials leads to power laws in the offspring-number distribution~\citep{Reed-Hughes2002,Newman2005,Niwa-etal2017}.

Recruitment is calculated by summing random offspring numbers,
where the number of summands (i.e. spawning population size) is also a random process.
Under the power-law model of offspring-number distribution,
the exponent $\alpha$ being less than two qualifies
the recruitment distribution as belonging to the L{\'e}vy-stable distribution regime
\citep{Levy1925,Khintchine-Levy1936,Levy1937},
so that the tail of the recruitment distribution follows a power law with the same exponent $\alpha$ and thus,
the recruitment distribution has diverging second moment (infinite variation).
One also expects that
the distribution of relative changes in recruitment (i.e. recruitment growth-rates) over a one-year interval is symmetric and L{\'e}vy-stable with the same exponent $\alpha$.

On the other hand,
the match-mismatch hypothesis \citep{Cushing1990} has been used to describe climate effects on the interannual variability in marine fish recruitment and thus, recruitment is primarily regulated by abiotic (i.e. density-independent) factors.
If individual reproductive outputs (or family sizes) have a small dispersion (relative to year-to-year variation in population sizes, or in carrying capacity of environment),
the reproductive variations in birth and death process lead to negligible aggregate fluctuations in a large population.
Consequently, environmentally driven fluctuations in (spawning) population sizes
appear to explain macroscopic (aggregate) fluctuations.
So that individual reproductive variations cannot be extracted from population-level measurements (e.g. interannual recruitment variability).

I show that,
in spite of strong fluctuations in the population size (carrying capacity),
the above argument on the (negligible) aggregate effect breaks down if the distribution of family sizes is heavy-tailed.
This paper discusses how, in a world with power-law offspring-number distribution (with exponent $1<\alpha<2$), idiosyncratic individual-level fluctuations (i.e. differences in individual reproductive successes, or variances of family sizes) aggregate up to non-trivial aggregate fluctuations in a large population under a randomly varying environment.
The idiosyncratic reproductive variability is quantitatively large enough to matter at the population level.
In this instance,
the distribution of recruitment growth-rates is not symmetric, but is instead totally asymmetric (skewed to the right) and heavy-tailed with the same exponent $\alpha$.
I contend that aggregate fluctuations come in tail part from idiosyncratic reproductive behavior,
even though the fluctuations of the sums of individual reproductive outputs are dominated by population-wide factors such as climate and fisheries.
Of particular interest is to what extent the idiosyncratic reproductive variations are averaged out upon aggregation under population fluctuations.

The paper is organized as follows.
Section~\ref{sect:model} introduces a recruitment model.
Section~\ref{sect:observations} reports some stylized facts regarding recruitment variations in commercially exploited fish stocks.
Section~\ref{sect:theory} presents model calculations in a stable-distribution framework based on random sums of random variables, which connect all those observations.
Then, Section~\ref{sect:simulation} numerically deals with nonvanishing aggregate fluctuations.
Finally, Section~\ref{sect:conclusion} concludes.

\section{Recruitment model}\label{sect:model}
The (spawning) population size is the number of reproductively-successful individuals in a generation.
Their offspring numbers are independent copies of random variable $X$ with asymptotic probability distribution
\begin{equation}\label{eqn:reproductive-variability}
 \mathrm{Pr}(X\geq x)\simeq (x_0/x)^{\alpha},
\end{equation}
which, with $1<\alpha<2$ and $x_0>0$, decays slowly as $x\to\infty$.
The offspring-number distribution has finite mean (assumed greater than one) and infinite variance.
Define $N_t$, the size of the population at time $t$, to be random variables with finite mean $\expval{N}$.
Angled brackets denote mean over all possible realizations.
The recruitment is the total number of offspring entering the (potentially reproductive) population.
Let $R_t$ denote the number of recruits to the population at time $t+1$ in years (or generations),
\begin{equation*}
 R_t=\sum_{n=1}^{N_t} X_n,
\end{equation*}
where $X_n$ is
the number of offspring of the $n$-th individual in year $t$ with the distribution in Equation~\eqref{eqn:reproductive-variability} being identical for all $t$
(i.e. the number of summands is independent of the summands $X_n$).
The offspring generation is constituted by sampling $N_{t+1}$ out of $R_t$ potential offspring.
The average recruitment is given by
\begin{equation*}
 \expval{R}=\expval{X}\expval{N}
\end{equation*}
with mean offspring number $\expval{X}>1$.
The amplitude of fluctuations in $N_t$ is given by
\begin{equation*}
 \Sigma_N^2 = \frac{\sum_{t=1}^T (N_t-\expval{N})^2}{T},
\end{equation*}
where $T$ is the number of trials (generations) of a random process.
The amplitude of fluctuations in $R_t$ is given by
\begin{equation*}
 \Sigma_R^2=\frac{\sum_{t=1}^T\qty(R_t-\expval{R})^2}{T}.
\end{equation*}

Temporal variation is considered by examining the relative (or absolute) variation of the recruitment, instead of the change in the logarithm of the recruitment.
Let $\eta_t$ be the relative change (i.e. the percentage growth rate) in recruitment over a year,
\begin{equation*}
 \eta_t = \frac{R_{t+1}-R_t}{R_t},
\end{equation*}
where $R_{t+1}-R_t$ is the increment (absolute difference) of recruitment between two successive years (or generations).
It is assumed that $\expval{\eta}=0$ and $\expval{R_{t+1}-R_t}=0$.
The amplitude of fluctuations in $\eta_t$ is given by
\begin{equation*}
 \Sigma_{\eta}^2 = \frac{\sum_{t=1}^{T-1}\eta_t^2}{T-1}.
\end{equation*}

When there are several stocks (subsystems with respect to the habitat of the species) for which one observes year-class recruitment,
the difference between stocks with smaller and greater mean recruitment comes from the different mean number of spawners.
One is interested in how the sample standard deviation changes, across stocks,
with the value of the mean annual recruitment to individual stocks.
One then calculates the $\expval{R^i}$ and $\Sigma_{R^i}$
(where the superscript $i$ denotes the stock identifier), and compares the couplings across stocks
to investigate whether stocks with larger mean recruitment exhibit larger fluctuations.
The relationship will be extracted as
\begin{equation}\label{eqn:FS-R}
 \Sigma_{R^i}\propto\expval{R^i}^b
\end{equation}
with scaling exponent $b$ being constant (and with the same prefactor) for all stocks.
This type of scaling relationship is called (a kind of) \citetapos{Taylor61} law or fluctuation scaling \citep{Eisler2008}.
The scaling law states that the fluctuations (the sample standard deviation of each stock) can be represented by a power-law function of the corresponding average.
The scaling exponent $b$ is usually between 1/2 and~1
\citep{Eisler2008}.
Then, the standard deviation in recruitment growth-rate $\eta^i$ should decay as the $(b-1)$ power of average recruitment,
\begin{equation}\label{eqn:FS-eta}
 \Sigma_{\eta^i}\propto\expval{R^i}^{b-1}.
\end{equation}
Model calculations employ a stable-distribution formalism based on random sums of random variables.
In this context the emergence of fluctuation scaling is equivalent to some corresponding limit theorems.

\section{Statistical analysis}\label{sect:observations}
This section provides insight in understanding the empirical properties of fish recruitment variability.

\subsection{Data}
Time-series data \citep[same as][]{Niwa-arxiv11feb2022} were taken from International Council for the Exploration of the Sea, archived at
\url{http://www.ices.dk/advice/Pages/Latest-Advice.aspx}.
There were 72 fish stocks analyzed throughout the North Atlantic:
19 pelagic stocks (7 species:
capelin, herring, horse mackerel, mackerel, sardine, sprat and blue whiting),
50 demersal stocks (16 species:
white anglerfish, cod, haddock, hake, Greenland halibut, megrim, four-spot megrim, plaice, Norway pout, beaked redfish, golden redfish, saithe, sandeel, seabass, sole and whiting),
2 deep-water stocks (2 species:
ling and tusk)
and one crustacean stock (northern shrimp).
The lengths of these time-series varied from 18 to 72 years.

\subsection{Empirical results}
\begin{figure}[htb!]
 \centering
   \begin{tabular}{ll}
    \small{(a)} & \small{(b)}\\
    \includegraphics[height=.3\textwidth,bb=0 0 360 232]{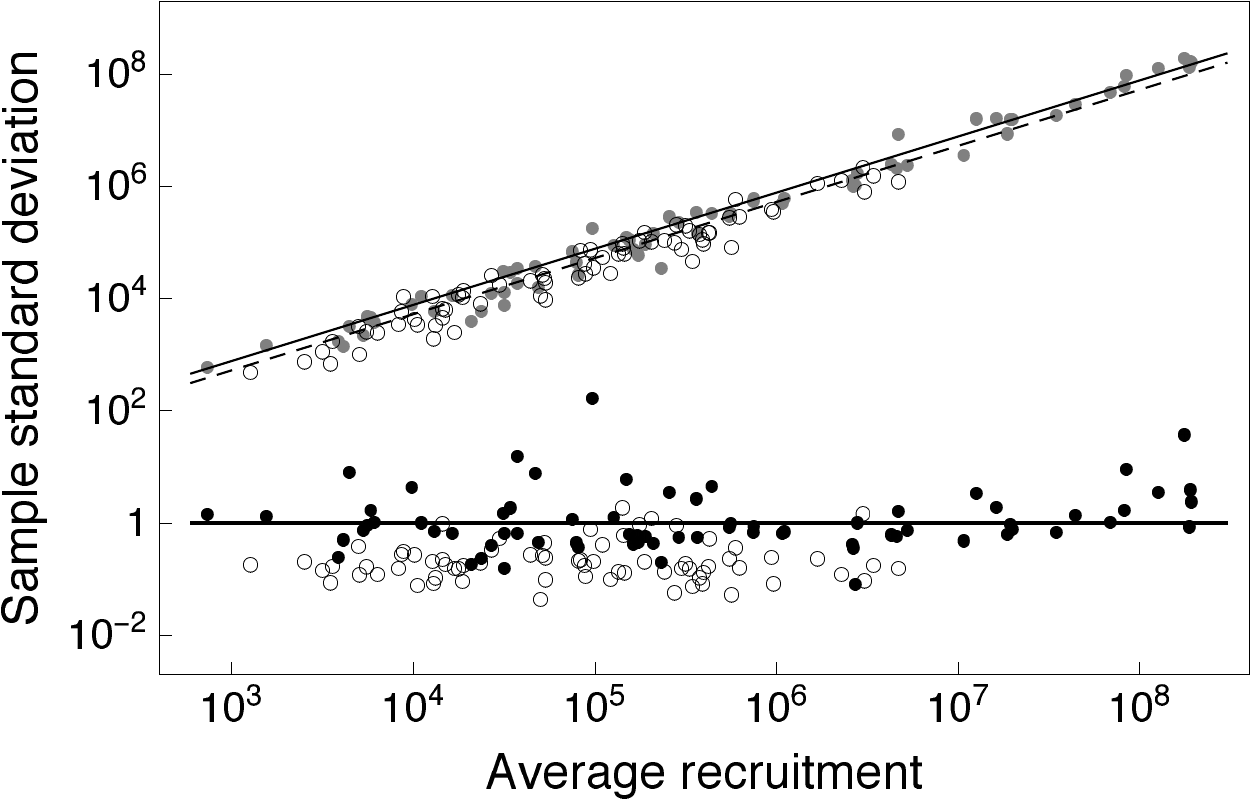}&
        \includegraphics[height=.3\textwidth,bb=0 0 360 230]{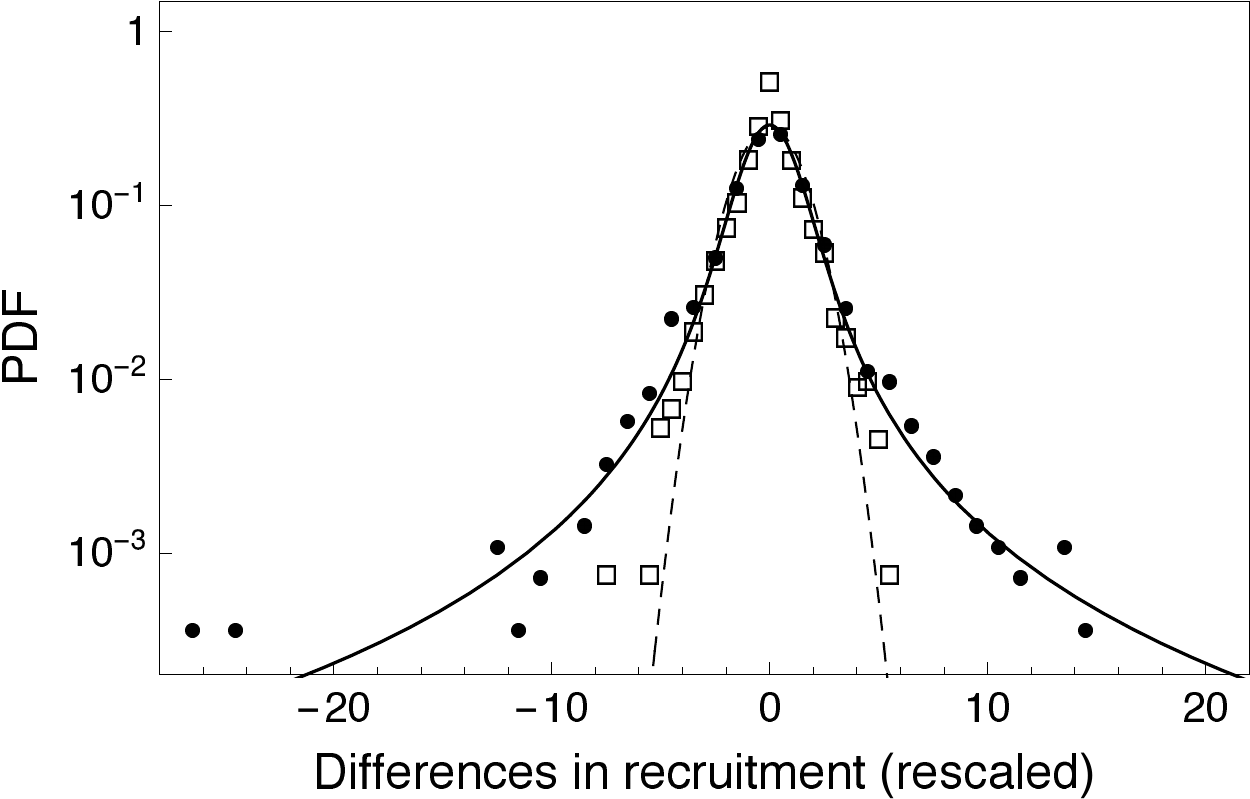}\\
    \small{(c)} & \small{(d)}\\
    \includegraphics[height=.3\textwidth,bb=0 0 360 222]{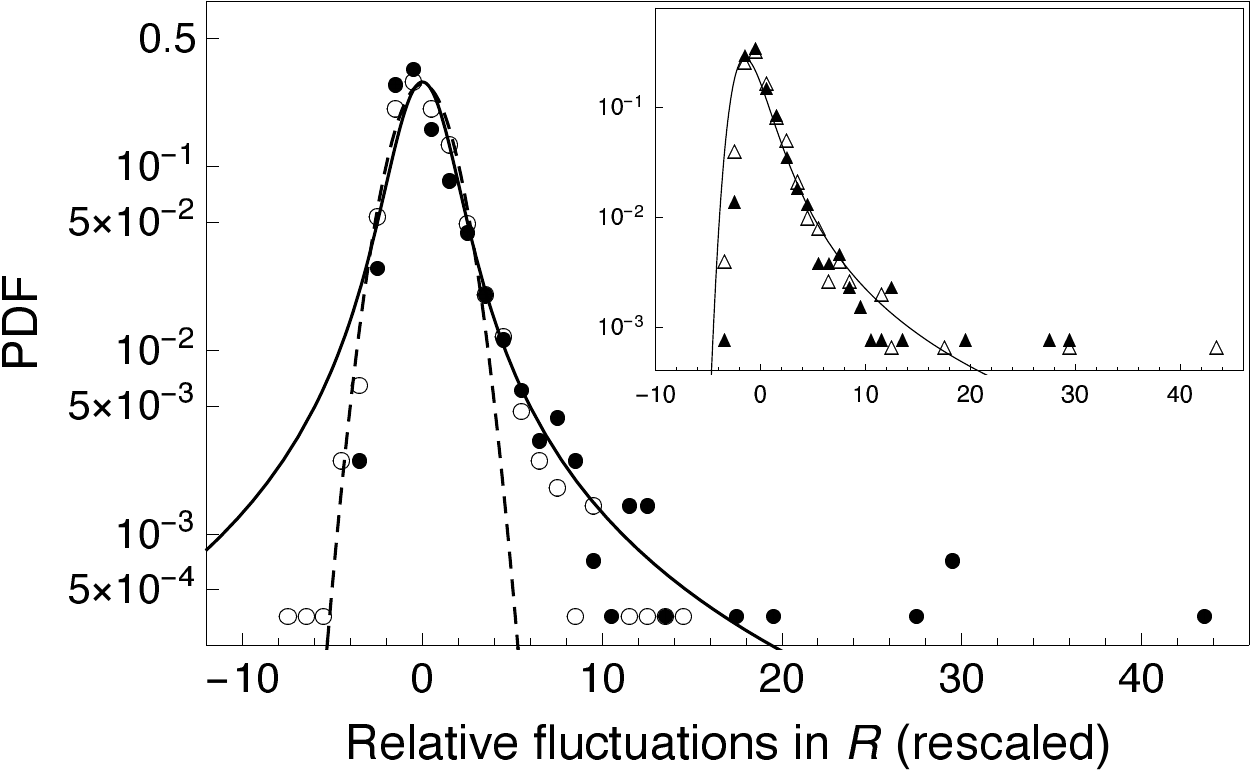}&
        \includegraphics[height=.3\textwidth,bb=0 0 360 235]{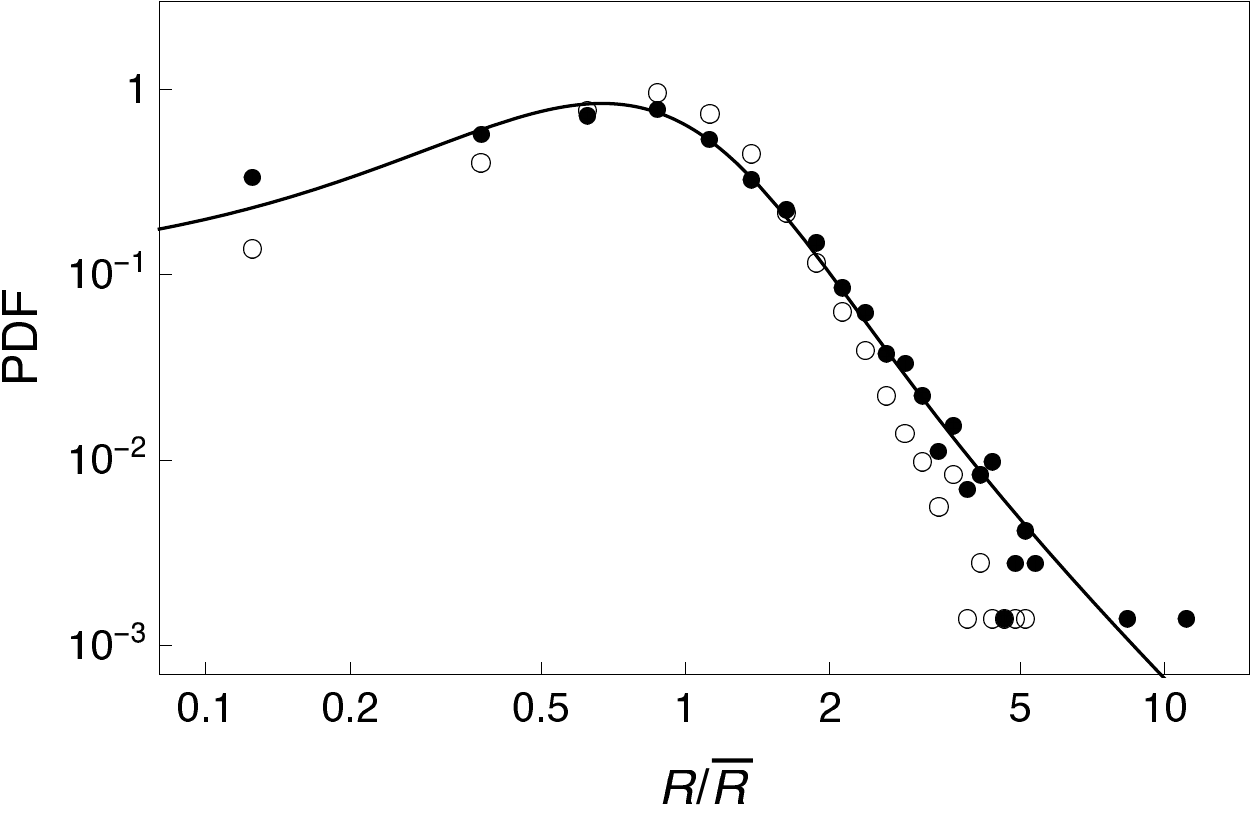}
      \end{tabular}
 \caption{\small
Statistical analysis of stock-recruitment data from the North Atlantic.
(a)~Fluctuation scaling for $\Sigma_R$ (upper solid gray circles) and $\Sigma_{\eta}$ (lower solid circles) among stocks ($R^i$ in units of $10^3$).
(b)~Leptokurtic distribution of successive differences in recruitment.
(c)~Asymmetric distributions of relative changes in recruitment (solid circles).
In the inset, open (resp. solid) triangles are plotted for autocorrelated (resp. non-autocorrelated) series of recruitment data.
(d)~Distributions of normalized recruitment (solid circles).
In these panels, open circles are for the spawning-stock biomass (in tonnes).
 }\label{fig:ices2017-TL}
\end{figure}

Figure~\ref{fig:ices2017-TL}a shows that the recruitment data (in thousand fish, upper solid gray circles) across 72 stocks (regardless of species) collapse onto the upper solid line
$\Sigma_R\propto\bar{R}$ with prefactor
\begin{equation*}
 \overline{\qty[\Sigma_{R^i}^2/\bar{R^i}^2]}^{1/2}=0.766
\end{equation*}
(average across stocks) over five orders of magnitude in $\bar{R}$,
where the overbar denotes sample mean.
The spawning biomass data (in tonnes, upper open circles) also exhibit linear scaling $\Sigma_{N}\propto\bar{N}$ with prefactor
\begin{equation*}
 \overline{\qty[\Sigma_{N^i}^2/\bar{N^i}^2]}^{1/2}=0.522
\end{equation*}
(dashed line).

Figure~\ref{fig:ices2017-TL}a also shows that
the sample standard deviation of relative changes in recruitment (lower solid black circles) hovers around $y=1$ (lower solid line), independent of $\bar{R^i}$.
The lower open circles plot the sample standard deviation of relative changes in spawning-stock biomass versus average abundance.

I now study the absolute differences between two consecutive values in recruitment series,
where the time series is detrended by subtracting the mean trend.
The successive differences are rescaled with the fluctuation width (i.e. scale factor) by stocks,
where the scale factor is estimated by the root-median-square successive differences of the (detrended) recruitment series
\citep{Takayasu-etal2014}.
The rescaled distributions with zero mean and unity width, when aggregated across 72 stocks, fit a symmetric L{\'e}vy-stable distribution with exponent $\hat\alpha=1.42$ (maximum likelihood estimate);
see Figure~\ref{fig:ices2017-TL}b (solid circles with solid line).
Further, using the standard deviation divided by $\sqrt{2}$,
when the absolute differences of recruitment are rescaled by stock and aggregated across stocks,
the shape of the distribution (open squares in Figure~\ref{fig:ices2017-TL}b) is more concentrated than the Gaussian distribution (with mean zero and standard deviation $\sqrt{2}$, dashed line), and the tails are heavier than it.

Figure~\ref{fig:ices2017-TL}c shows that
the relative changes in recruitment, when aggregated across 72 stocks, follow closely the maximally asymmetric L{\'e}vy-stable distribution with exponent $\hat\alpha$ (with the tail vanishing to the left),
where the data are rescaled with their root-median-square values by stock,
after detrended by subtracting the mean trend.
The solid line in the main panel is the symmetric L{\'e}vy-stable distribution with exponent $\hat\alpha$ (and with zero mean and unity width), and the dashed line is the Gaussian distribution with mean zero and standard deviation $\sqrt{2}$.
The relative changes in spawning-stock biomass, rescaled with their root-median-square values by stock (after detrended by subtracting the mean trend), also exhibit totally skewed or asymmetric character (open circles, with data aggregated across stocks), which is close to the rescaled distribution of recruitment growth-rates.
Temporal autocorrelation may act to increase or decrease the fluctuation width of growth rates.
Figure~\ref{fig:ices2017-TL}c (inset) shows that
for 36 (solid triangles) out of 72 recruitment series that have passed the Ljung-Box test at the 5\% level, which reasonably satisfy the assumption of no autocorrelation,
the recruitment growth-rates collapse onto those
for the rest (open triangles) of the recruitment series showing some evidences of autocorrelation,
where data are aggregated across stocks after rescaled by the fluctuation width.
The solid line in the inset is the maximally asymmetric L{\'e}vy-stable distribution with exponent $\hat\alpha$ (with zero mean and unity width).

Figure~\ref{fig:ices2017-TL}d shows that the normalized recruitment data $R_t^i/\bar{R^i}$ (solid circles, aggregated across stocks) collapse upon a universal curve (scaling function).
I verify the fitting of a L{\'e}vy-stable distribution (solid line).
Under the stable law with $\hat\alpha=1.42$,
the degree of asymmetry is estimated at one by maximum likelihood,
and the maximum likelihood estimate of location (resp. scale) is $1.16$ (resp. $0.332$).
Furthermore, the distribution of the normalized variables $R_t^i/\bar{R^i}$ is close to that for the $N_t^i/\bar{N^i}$ (open circles) in the central part.

\subsection{Stylized facts on recruitment fluctuations}
Analyzing fisheries stock-recruitment data from the North Atlantic, I have found the following results:
(i)~the linear dependence of the amplitude of recruitment fluctuations on recruitment level,
(ii)~the independence of the variation in relative recruitment changes from recruitment level,
(iii)~the leptokurtic character of (absolute) recruitment increments,
and (iv)~the totally asymmetric character of the distribution of relative recruitment changes.

The linear scaling of fluctuations in abundance of populations was reported for the North Sea fish community
\citep{Cobain2019}.
In this paper, I have found that
the per-recruit variability in recruitment is close to the per-capita variability in population abundance, and both of these are constant for all stocks.
The scaling exponent of $b=1$ as in Equation~\eqref{eqn:FS-R} indicates that the recruitment series $R_t^i$'s are rescaled with $\bar{R^i}$ (times a constant),
and the normalized observations have a universal distribution.
The scaling function is very close to a maximally asymmetric L{\'e}vy-stable distribution.

The asymmetric (skewed to the right) character of the distribution of relative changes has been repeatedly observed in a variety of animal (insect, bird, mammal and fish) populations
\citep{Keitt-Stanley98,Allen-etal01,Keitt-etal02,Niwa06-ecoinf,Niwa07-ICES,Lan2011}.
Another, much less studied property is the scaling relationship between the standard deviation and average of the growth-rate distribution
\citep{Keitt-etal02,Niwa06-ecoinf}.
In this paper, I have found that
the amplitude of fluctuations in recruitment growth-rate $\eta$ is approximately independent of the recruitment level,
which is a version of \citetapos{Gibrat31} law for variances.
The amplitude of fluctuations in year-to-year percentage population growth-rate is also approximately independent of population size.

A conventional approach uses the log-normal distribution of fish recruitment
\citep{Hilborn-Walters92}.
The ratio of two numbers, each drawn from a log-normal distribution, is also log-normally distributed, so the distribution of $R_{t+1}/R_t{\,}(=\eta_t+1)$ would be log-normal as well.
Further, the difference of two log-normal random variables should resemble a Gaussian distribution
\citep{Carmona-Durrleman2003}.
I have found that the distribution of (absolute) recruitment increments is not Gaussian but rather leptokurtic.
The representation of recruitment increments as symmetric L{\'e}vy-stable random variables may account for their leptokurtic character.
However, it has a drawback:
the relative fluctuations in recruitment have totally skewed properties, showing in fact they appear to be a maximally asymmetric L{\'e}vy-stable distribution.
These apparently contradictory aspects will be blended together into a consistent picture in the following section.
A statistical test for log-normality of the data is provided in \citet{Niwa-arxiv11feb2022}.

\section{Model calculations}\label{sect:theory}
\subsection{L{\'e}vy-stable fluctuations}
Consider the sum of $N$ independent random variables $X_1,\ldots,X_N$
with a power-law distribution in Equation~\eqref{eqn:reproductive-variability} of exponent $\alpha\in (1,2)$.
If the population size $N$ is fixed constant in an environment with constant carrying capacity,
the sum $R=\sum_{n=1}^N X_n$ has a mean $\expval{R}=\expval{X}N$,
and as $N\to\infty$,
the rescaled sum $(R-\expval{R})/N^{1/\alpha}$ has a limit distribution, i.e. a maximally asymmetric L{\'e}vy-stable law of exponent~$\alpha$.
The tail vanishes to the left, and the assumption $\expval{X}>1$ ensures that $R>N$ with sufficiently high probability.
Since $X_n^2$ follows a power-law distribution with exponent $\alpha/2<1$,
one has, as $N\to\infty$,
\begin{equation*}
 \qty(R-\expval{R})^2 \approx
  \sum_{n=1}^N\qty(X_n-\expval{X})^2
  \simeq x_0^2 C_{\alpha/2}N^{2/\alpha} w
\end{equation*}
with $C_{\alpha/2}=[\Gamma(1-\alpha/2)\cos(\pi\alpha/4)]^{2/\alpha}$,
where $w$ is a random variable with maximally asymmetric L{\'e}vy-stable distribution with exponent $\alpha/2$ and width (scale factor) of unity
\citep{Gabaix2011}.
The distribution of $w$ does not depend on $N$.
The recruitment growth-rate follows
\begin{equation*}
 \eta^2\simeq
  2^{2/\alpha} \qty(x_0/\expval{X})^2 C_{\alpha/2} N^{2(1/\alpha-1)}w.
\end{equation*}
Accordingly, the recruitment variability scales as
\begin{align}
 \label{eqn:idiosyncratic-fluctuations-R}
 \Sigma_R &\propto \expval{R}^{1/\alpha} T^{1/\alpha-1/2}\\
 \label{eqn:idiosyncratic-fluctuations-eta}
 \Sigma_{\eta} &\propto \expval{R}^{1/\alpha-1} T^{1/\alpha-1/2}
\end{align}
with exponent $b=1/\alpha$ as in Equations~\eqref{eqn:FS-R} and~\eqref{eqn:FS-eta},
where $T{\,}(\gg 1)$ is the number of trials
\citep{Bouchaud-Georges90,Newman2005}.
Note that the variances $\expval{(R-\expval{R})^2}$ and $\expval{\eta^2}$ are infinite.

Define $X_{1,N}\geq X_{2,N}\geq\cdots\geq X_{N,N}$ by ranking in decreasing order the values encountered among the $N$ terms of the sum $R$.
When $1<\alpha<2$, one has, for large $N$,
\begin{align*}
 \expval{X_{1,N}}&=\frac{x_0 N!{\,}\Gamma(1-1/\alpha)}{\Gamma(N+1-1/\alpha)}
 \approx x_0\Gamma(1-1/\alpha)N^{1/\alpha}\\
 \expval{X_{2,N}}&=\frac{\alpha-1}{\alpha}\expval{X_{1,N}}
\end{align*}
and while $X_{1,N}$ has an infinite second moment (infinite variation), one has
\begin{align*}
 \expval{X_{2,N}^2}&=\frac{x_0^2 N!{\,}\Gamma(2-2/\alpha)}{\Gamma(N+1-2/\alpha)}
 \approx x_0^2\Gamma(2-2/\alpha)N^{2/\alpha}\\
 \expval{X_{1,N}X_{2,N}}&=\frac{\alpha}{\alpha-1}\expval{X_{2,N}^2}
\end{align*}
\citep{ZKS05},
where the Euler product formula for the gamma function is used.
Importantly, all but the largest order statistics have finite second moment.
Therefore, the sum $R_{2,N}{\,}(=\sum_{k=2}^N X_{k,N})$ of the $(N-1)$ lower order statistics converges
to a Gaussian-distributed random variable with first two moments given by
\begin{equation*}
 \expval{R_{2,N}}\approx
 \frac{x_0\alpha}{\alpha-1}\qty(N-\Gamma(2-1/\alpha)N^{1/\alpha})
 \approx\expval{R}
\end{equation*}
and
\begin{equation*}
 \expval{R_{2,N}^2}=
  \expval{\sum_{k=2}^N X_{k,N}^2}
  +\expval{\sum_{k\neq\ell} X_{k,N}X_{\ell,N}}
  \sim N^{2/\alpha},
\end{equation*}
where
\begin{equation*}
 \expval{\sum_{k=2}^N X_{k,N}^2}\approx
  \frac{x_0^2\alpha}{\alpha-2}
  \qty(N-\Gamma(2-2/\alpha)N^{2/\alpha})
  \approx\frac{\alpha}{2-\alpha}\expval{X_{2,N}^2},
\end{equation*}
and where
\begin{align*}
 \expval{X_{2,N}X_{3,N}} &=
  \frac{2\alpha}{2\alpha-1}\expval{X_{3,N}^2}\\
 \expval{X_{2,N}X_{4,N}} &=
  \frac{3!\alpha^2}{(2\alpha-1)(3\alpha-1)}\expval{X_{4,N}^2}
\end{align*}
etc.
Accordingly, while the sum $R$ is dominated by its largest term $X_{1,N}$,
the sample-to-sample variations in recruitment ($\Sigma_R$) and in growth rates ($\Sigma_{\eta}$) are sensitive to central observations in the vast majority
and insensitive to the rare tail events.

The recruitment $R$ follows a power-law distribution with exponent $\alpha$
in the tail
\begin{equation*}
 R-\expval{R}\gtrsim N^{1/\alpha},
\end{equation*}
i.e. $R\gtrsim\expval{X}N$ for large $N$,
so the fraction of the total recruits, $(R-\expval{R})/\expval{R}\sim N^{1/\alpha-1}$, can be linked to one parent.
The recruitment growth-rates are symmetrically and L{\'e}vy-stable distributed,
and follow a power law asymptotically with exponent $\alpha$ in the tails
\begin{equation*}
 \abs{\eta}\gtrsim N^{1/\alpha-1},
\end{equation*}
because the size of the largest family $X_{1,N}$ scales as $N^{1/\alpha}$.

\subsection{Beyond the L{\'e}vy-stable fluctuations}
To incorporate externally induced fluctuations, let us allow $N$ to vary independently from one year to the other.
The $R_t$'s are independent.
Recruitment variability contains contributions as a result of individual reproductive variations and population-size (or carrying-capacity) fluctuations,
both of which are sources of fluctuations in output productivity.
It then follows that
\begin{equation*}
 \qty(R_t-\expval{R})^2=
 \qty(R_t-\expval{X}N_t)^2
 +\expval{X}^2\qty(N_t-\expval{N})^2
 +2\expval{X}\qty(R_t-\expval{X}N_t)\qty(N_t-\expval{N}),
\end{equation*}
yielding
\begin{equation*}
 \Sigma_R^2
  \simeq
  x_0^2 C_{\alpha/2}\expval{N}^{2/\alpha} T^{2/\alpha-1} w
  +\expval{X}^2\Sigma_N^2
\end{equation*}
for large $\sum_{t=1}^T N_t\approx\expval{N}T$.
Assume $\Sigma_N/\expval{N}\ll 1$, so that $\Sigma_R/\expval{R}\ll 1$.
Then, one has
\begin{equation*}
 \Sigma_{\eta}^2\approx 2\Sigma_R^2/\expval{R}^2.
\end{equation*}

When the amplitude of fluctuations $\Sigma_N$ is typically less than $\expval{N}^{1/\alpha}$, i.e. the ratio ${\Sigma_N}/{\expval{N}^{1/\alpha}}$ approaches 0 as $\expval{N}$ gets large,
one recovers Equations~\eqref{eqn:idiosyncratic-fluctuations-R} and~\eqref{eqn:idiosyncratic-fluctuations-eta}.

When the $\Sigma_N$ exceeds the threshold, i.e.
the ratio ${\Sigma_N}/{\expval{N}^{1/\alpha}}$ goes to infinity as $\expval{N}\to\infty$
with $\Sigma_N/\expval{N}$ kept finite ($\ll 1$),
one typically has
\begin{equation*}
 \Sigma_R\approx\expval{X}\Sigma_N,
\end{equation*}
which leads to
\begin{equation*}
 \Sigma_R/\expval{R}\approx\Sigma_N/\expval{N}.
\end{equation*}
The typical relative change in recruitment over a year
is then estimated by
\begin{equation}\label{eqn:fluctuation-eta}
 \Sigma_{\eta}\approx\sqrt{2}{\,}{\Sigma_{N}}/{\expval{N}}.
\end{equation}
Therefore,
under environmentally driven population fluctuations with $\Sigma_N\propto\expval{N}$,
the recruitment variability scales as
\begin{align*}
 \Sigma_R &\propto \expval{R}\\
 \Sigma_{\eta} &\sim 1
\end{align*}
with exponent $b=1$ as in Equations~\eqref{eqn:FS-R} and~\eqref{eqn:FS-eta},
where the prefactors in the scaling relations may depend on $T$ through $\Sigma_N$.

A generic representation of the recruitment dynamics is given by
\begin{equation}\label{eqn:transform-R}
 R_t=x_0 C_{\alpha}N_t^{1/\alpha}z_t+\expval{X}N_t
\end{equation}
with $z_t$ being distributed as the maximally asymmetric L{\'e}vy-stable law of exponent $\alpha$ with zero mean and unity width,
where $C_{\alpha}=[\Gamma(1-\alpha)\cos(\pi\alpha/2)]^{1/\alpha}$.
The variable $z_t$ is independent of $N_t$.
If $\Sigma_N/\expval{N}^{1/\alpha}\to 0$ as $\expval{N}\to\infty$,
one has
\begin{equation}\label{eqn:pdf-R0}
 R_t\simeq x_0 C_{\alpha}\expval{N}^{1/\alpha}z_t+\expval{X}\expval{N},
\end{equation}
and consequently,
\begin{equation*}
 \eta_t\simeq x_0 C_{\alpha}
  \frac{z_{t+1}-z_t}{\expval{X}\expval{N}^{1-1/\alpha}},
\end{equation*}
which follows a symmetric L{\'e}vy-stable distribution with exponent $\alpha$.
If $\Sigma_N/\expval{N}^{1/\alpha}\to\infty$ as $\expval{N}\to\infty$,
one has
\begin{equation}\label{eqn:pdf-R}
 R_t\simeq\expval{X}N_t,
\end{equation}
and consequently,
\begin{equation*}
 \eta_t\simeq\frac{N_{t+1}-N_t}{N_t}.
\end{equation*}

As the $\Sigma_N$ increases with increasing environmental disturbances,
the fluctuation width of $\eta_t$ broadens, asymptotically to give Equation~\eqref{eqn:fluctuation-eta}.
When
\begin{equation}\label{eqn:tail-eta}
 \abs{\eta_t}\gtrsim\sqrt{2}{\,}\Sigma_{N}/\expval{N}
\end{equation}
with scaling $\Sigma_N\propto\expval{N}$,
idiosyncratic variations do not die out in the aggregate,
and $\eta_t$'s follow a power-law distribution with the same exponent $\alpha$ as the offspring-number distribution.
Therefore, the growth-rate distribution is maximally asymmetric for large $\expval{N}$, since the tail vanishes to the left.
Besides the recruitment growth-rates,
the recruitment distribution also has a putative power-law regime with exponent $\alpha$ in the tail
\begin{equation}\label{eqn:tail-R}
 R_t-\expval{R}\gtrsim\Sigma_R,
\end{equation}
i.e. $R_t\gtrsim\expval{R}$.
So the outer tails still reflect the idiosyncratic reproductive fluctuations.
Contrariwise, the recruitment movements mimic the population-size (carrying-capacity) fluctuations in the central regions
\begin{align}
 \label{eqn:bulk-R}
 R_t &\lesssim\expval{R}+\Sigma_R\\
 \label{eqn:bulk-eta}
 \abs{\eta_t} &\lesssim\Sigma_{\eta}
\end{align}
where individual offspring-number fluctuations have a negligible aggregate effect in the limit of large $\expval{N}$.
The right-hand sides of Equations~(\ref{eqn:tail-eta}--\ref{eqn:bulk-eta})
are evaluated in the limit of $T\to 1$, if the $\Sigma_N$ depends on $T$.

\begin{remark*}
When environmental stochasticity imposes strong fluctuations in $N$,
if $\Sigma_N\propto\expval{N}$,
the linear-scaling behavior, $\Sigma_R\propto\expval{R}$, emerges
\citep[see also][]{Anderson82,deMenezes-Barabasi92,Niwa-arxiv11feb2022}.
If the per-capita variability in population abundance ($\Sigma_{N^i}/\expval{N^i}$) is constant across stocks,
it is assumed that
the normalized abundances, $\nu_t=N_t^i/\expval{N^i}$, are identically distributed across stocks
\citep{Eisler2008}.
If this is true, one has
\begin{equation*}
 \Sigma_{N^i}=\Sigma_{\nu}\expval{N^i}
\end{equation*}
with $\Sigma_{\nu}^2=\overline{[\Sigma_{N^i}^2/\expval{N^i}^2]}$
(average across stocks).
Further, from Equation~\eqref{eqn:pdf-R} as $N\to\infty$,
the normalized recruitment $R_t^i/\expval{R^i}$ is equal (in distribution) to the universal random variable $\nu_t$,
which gives rise to a linear relationship
\begin{equation}\label{eqn:linear-scaling}
 \Sigma_{R^i}\approx\Sigma_{\nu}\expval{R^i}.
\end{equation}
Even though the tail exponents of the offspring distributions are stock-specific, Equation~\eqref{eqn:linear-scaling} holds.
In this case, the normalized variables $R_t^i/\expval{R^i}$ are almost identically distributed on the central part across stocks.
\end{remark*}

\section{Simulation}\label{sect:simulation}
Consider the environmental (carrying-capacity) fluctuations as embedded in a confining well or barrier, ensuring mean-reverting behavior, without having to rely on details of regulation in the system.
$N_t$ individuals in year $t$ are sampled (selected for reproduction) out of $R_{t-1}$ extant individuals in year $t-1$.
A random variable $N_t$ is drawn according to a truncated (maximally asymmetric) L{\'e}vy-stable distribution bounded below at $R_{t-1}$,
where the distribution has a scale factor $\varsigma_N$ and a location parameter $\mu_N>0$.
The population size $N_t$ is given by
\begin{equation}\label{eqn:transform-N}
 N_t = \varsigma_N z'_t +\mu_N,
\end{equation}
where $z'_t$ is a random draw from the truncated (maximally asymmetric) L{\'e}vy-stable law of exponent $\alpha'\in (1,2]$ with support $(-\mu_N/\varsigma_N,(R_{t-1}-\mu_N)/\varsigma_N]$.
The recruitment $R_t$ is then given by Equation~\eqref{eqn:transform-R}.
Assume $\varsigma_N/\mu_N\ll 1$.
If $R_{t-1}-\mu_N>\varsigma_N$,
since the region $\abs{N_t-\mu_N}\gtrsim\varsigma_N$ is typically not sampled,
then one typically has
$R_t-\mu_N>\varsigma_N$ for a large population with $\expval{X}>1$.
So that the stock-recruitment process is insensitive to the confinement.
Consequently, one has $\expval{N}\approx\mu_N$, $\expval{R}\approx\expval{X}\mu_N$,
and $\Sigma_N\approx\varsigma_N T^{1/\alpha'-1/2}$
with $T{\,}(\gg 1)$ denoting the length of the stochastic process.

By setting $\alpha=\alpha'=1.42$,
and using a population of $\mu_N=10^5$ to $10^{10.5}$,
and three different $\varsigma_N$ values chosen to give
$\varsigma_N=\mu_N^{1/2}$, $\mu_N^{1/\alpha}$ and $0.1\mu_N$,
I simulated the stock-recruitment process for $10^5$ generations for each choice of parameter values, where $\expval{X}=\alpha/(\alpha-1)$ with $x_0=1$ in Equation~\eqref{eqn:reproductive-variability}.

\begin{figure}[htb!]
 \centering
 \begin{tabular}{ll}
  \small{(a)} & \small{(b)}\\
  \includegraphics[height=.3\textwidth,bb=0 0 360 234]{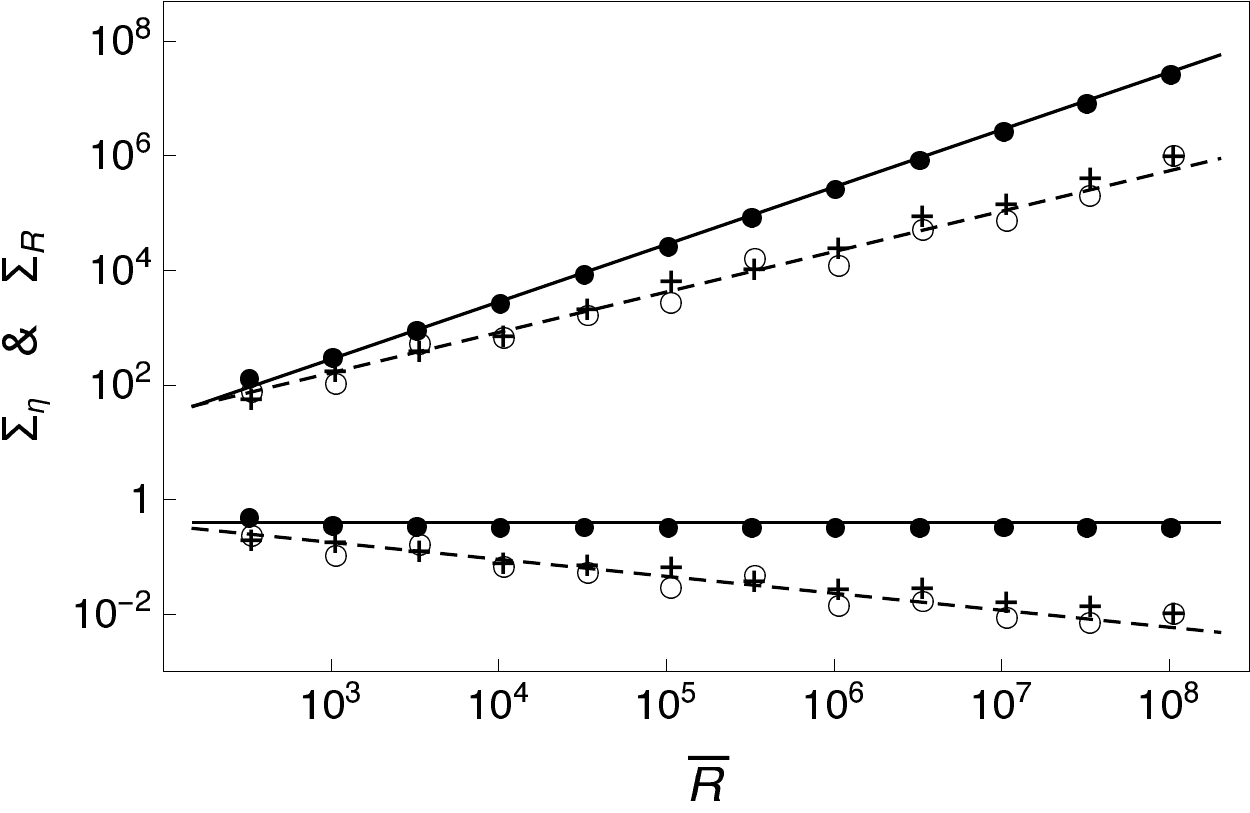}&
  \includegraphics[height=.3\textwidth,bb=0 0 360 230]{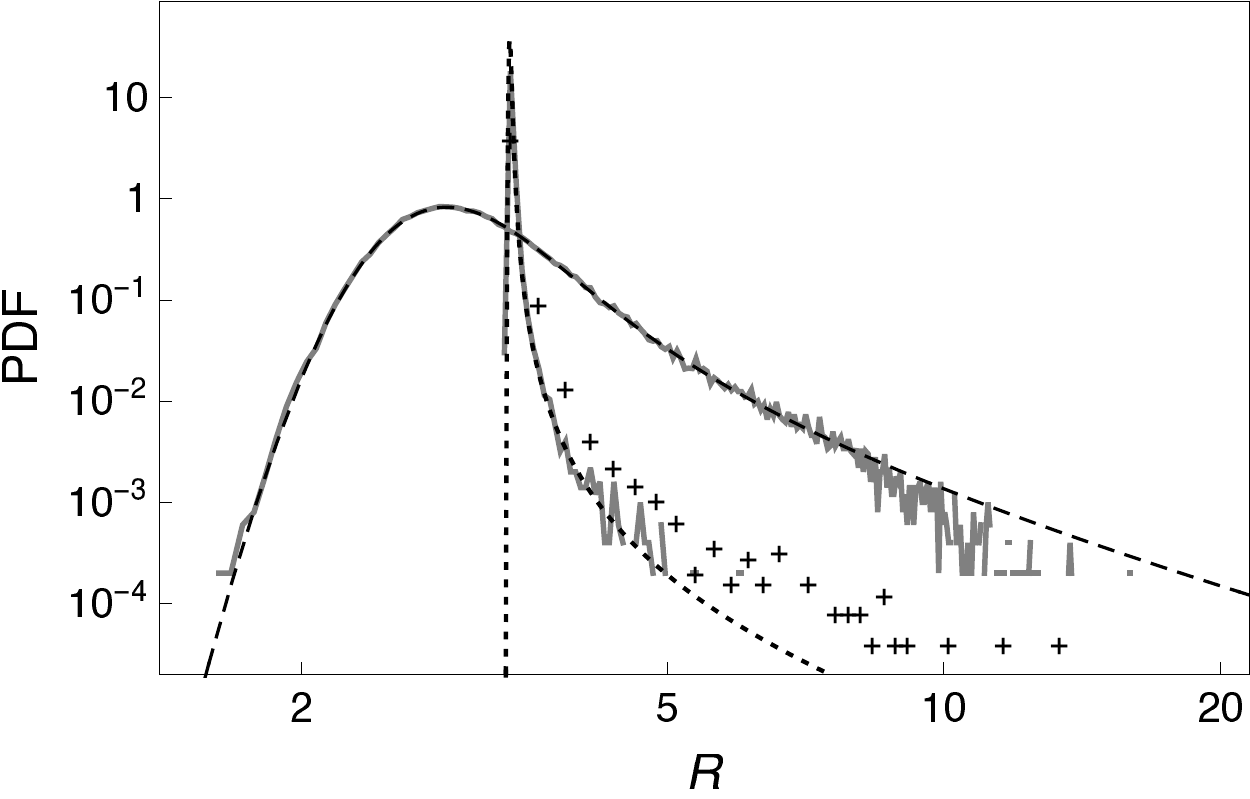}\\
  \small{(c)} &\\
  \includegraphics[height=.3\textwidth,bb=0 0 360 230]{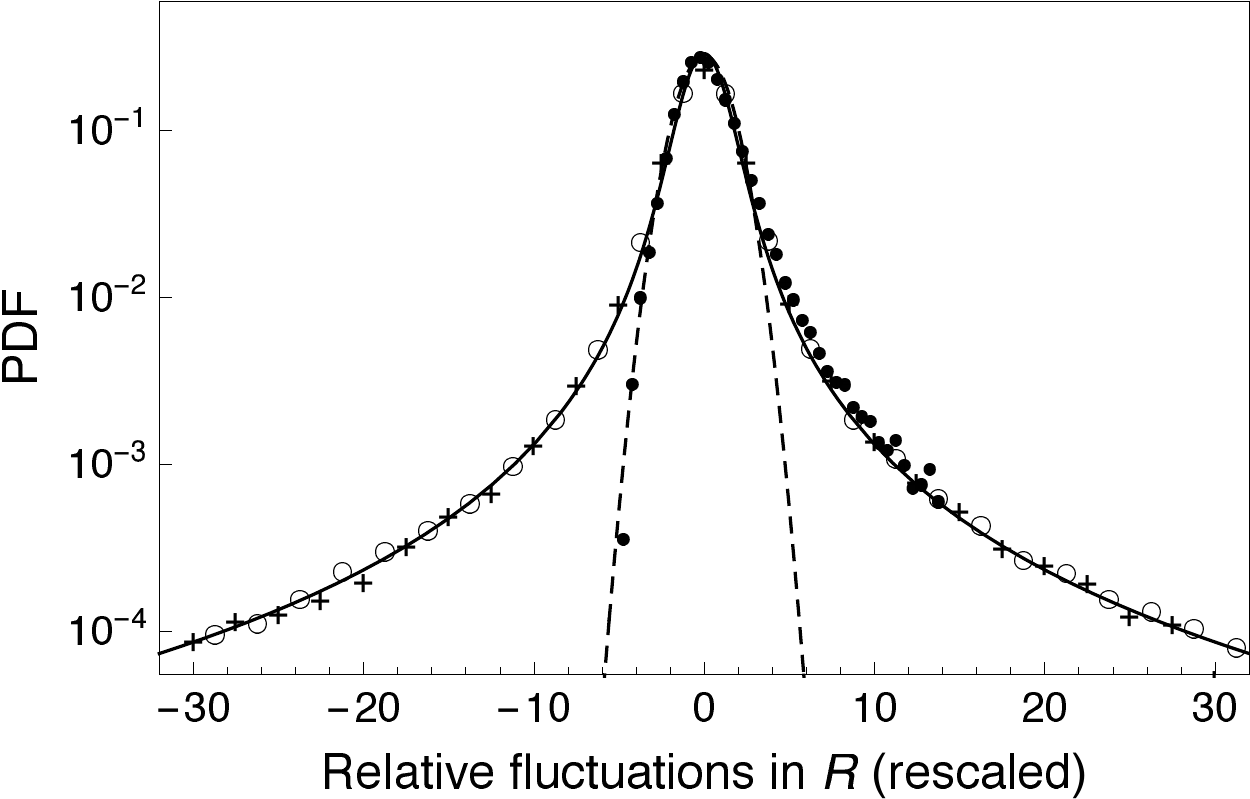}&\\
  \end{tabular}
 \caption{\small
L{\'e}vy-stable model and beyond.
Plots are generated by setting $\varsigma_N=\mu_N^{1/2}{\;(\circ)}$, $\mu_N^{1/\alpha}{\;(+)}$, and $0.1\mu_N{\;(\bullet)}$.
 (a)~Fluctuation scaling of $R$ (upper points) and $\eta$ (lower points).
Averages are taken over simulations ($R$ in units of $10^3$).
 (b)~Recruitment distributions (plotted on log-log scale, with $R$ in unit of $10^8$), generated by setting $\mu_N=10^8$ and $\expval{X}=3.38$.
 (c)~Rescaled distributions of recruitment growth-rates.
}\label{fig:simulation}
\end{figure}

The case of $\varsigma_N=\mu_N^{1/2}$
shows in Figure~\ref{fig:simulation}a that $\Sigma_R\propto\bar{R}^{1/\alpha}$
(upper open circles with dashed line) with constant prefactor
$\overline{[\Sigma_R/\bar{R}^{1/\alpha}]}=1.26$ (average across simulation sets),
and $\Sigma_{\eta}\propto\bar{R}^{1/\alpha-1}$ (lower open circles with dashed line) with
$\overline{[\Sigma_{\eta}/\bar{R}^{1/\alpha-1}]}=1.39$.
The average $\bar{R}$ is calculated from each simulation run.
The case of $\varsigma_N=\mu_N^{1/\alpha}$
shows in Figure~\ref{fig:simulation}a that the $\Sigma_R$ is still proportional to $\bar{R}^{1/\alpha}$ (upper plus signs),
and $\Sigma_{\eta}\propto\bar{R}^{1/\alpha-1}$ (lower plus signs).
The points for $R$ (resp. $\eta$) collapse onto the upper (resp. lower) dashed line.
The case of $\varsigma_N=0.1\mu_N$
shows in Figure~\ref{fig:simulation}a that $\Sigma_R\propto\bar{R}$ (upper solid circles with solid line) with $\overline{[\Sigma_R/\bar{R}]}=0.283$,
and the $\Sigma_{\eta}$ is independent of $\bar{R}$ (lower solid circles) with $\overline{[\Sigma_{\eta}]}=0.357$.
The (lower) solid horizontal line is at $\sqrt{2}{\,}\overline{[\Sigma_R/\bar{R}]}=0.400$.

As to the recruitment and growth-rate distributions,
Figure~\ref{fig:simulation}b and~c confirm that as $\expval{N}$ gets large,
while, when $\Sigma_N/\expval{N}^{1/\alpha}\to 0$, the fluctuations in recruitment and in growth rates are insensitive to external forces,
when $\Sigma_N/\expval{N}^{1/\alpha}\to\infty$,
the recruitment mimics the population fluctuations except for $R\gtrsim\expval{X}(\mu_N+\varsigma_N)$ and for $\eta\gtrsim\sqrt{2}{\,}\varsigma_N/\mu_N$.

Figure~\ref{fig:simulation}b shows the recruitment distributions for the population of $\mu_N=10^8$.
Equation~\eqref{eqn:pdf-R0} (dotted line with $\varsigma_N=\mu_N^{1/2}$) and
Equation~\eqref{eqn:pdf-R} with $N_t$ as in Equation~\eqref{eqn:transform-N}
(dashed line with $\varsigma_N=0.1\mu_N$) agree with the results (noisy gray curves) from the simulations.
The dotted line represents the maximally asymmetric L{\'e}vy-stable distribution ($\alpha=1.42$) with mean of $\expval{X}\mu_N$ and width of $x_0 C_{\alpha}\mu_N^{1/\alpha}$.
The dashed line represents the maximally asymmetric L{\'e}vy-stable distribution ($\alpha=1.42$) with mean of $\expval{X}\mu_N$ and width of $\expval{X}\varsigma_N=0.1\expval{X}\mu_N$.

Figure~\ref{fig:simulation}c shows the recruitment growth-rate distributions for the population of $\mu_N=10^8$.
The solid line represents the symmetric L{\'e}vy-stable distribution (with exponent $\alpha=1.42$, mean of zero and width of unity),
and the dashed line represents the Gaussian distribution (with mean zero and standard deviation $\sqrt{2}$).
For the case $\varsigma_N=\mu_N^{1/2}$ (open circles), the $\eta_t$'s are rescaled with $2^{1/\alpha}C_{\alpha}\mu_N^{1/\alpha-1}/\expval{X}$.
For the cases $\varsigma_N=0.1\mu_N$ (solid circles) and $\varsigma_N=\mu_N^{1/\alpha}$ (plus signs),
the $\eta_t$'s are rescaled with
their root-median-square values.

\begin{figure}[htb!]
 \centering
 \begin{tabular}{ll}
  \small{(a)} & \small{(b)}\\
  \includegraphics[height=.3\textwidth,bb=0 0 360 236]{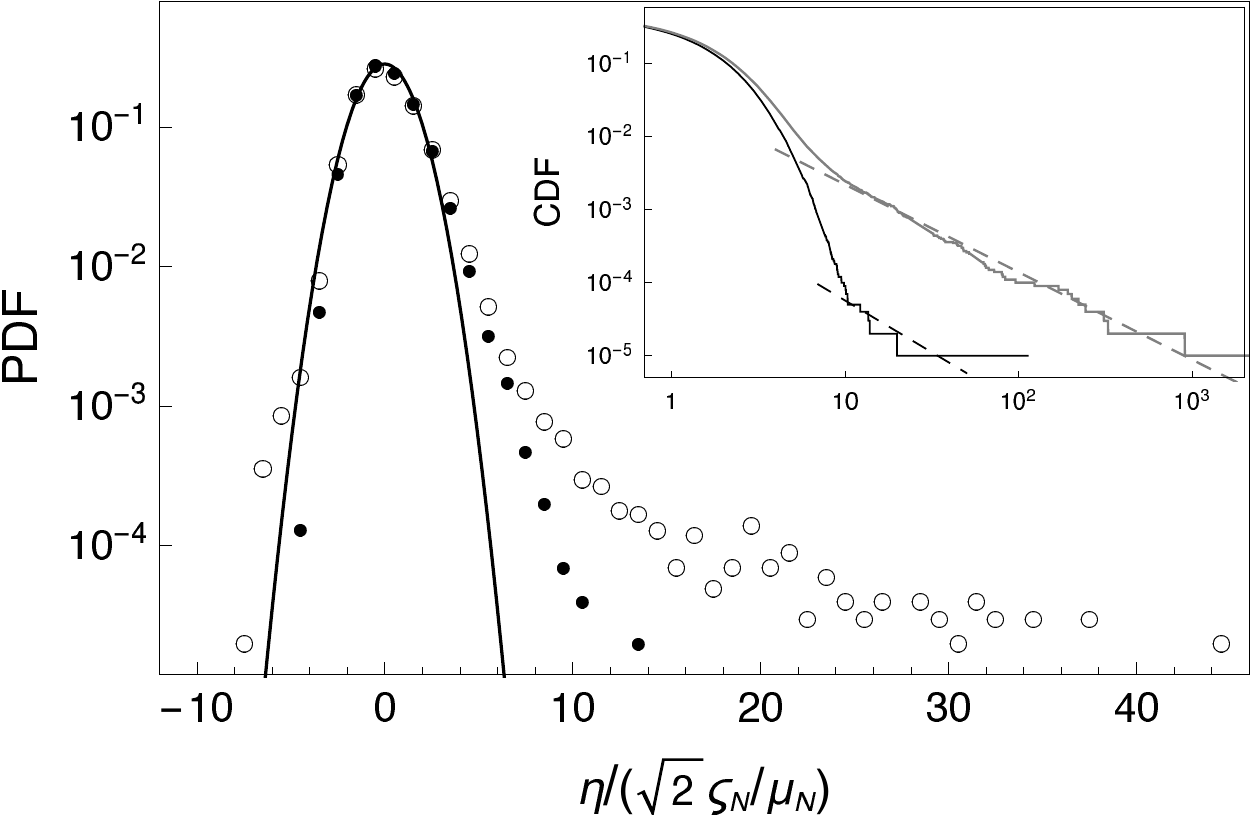}&
      \includegraphics[height=.3\textwidth,bb=0 0 360 232]{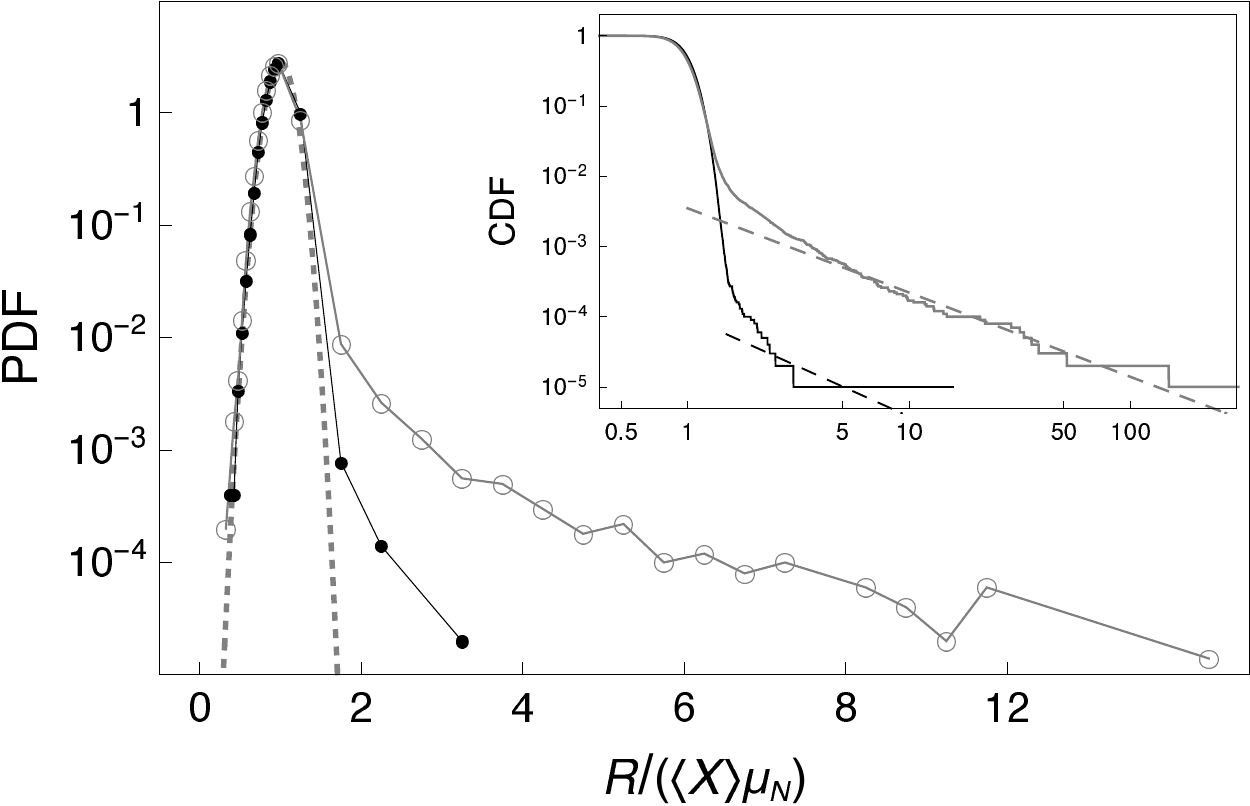}
 \end{tabular}
 \caption{\small
Aggregate effects of idiosyncratic variations.
 Plots are generated by setting $\alpha'=2$ and two different values of $\alpha=1.42$ (solid circles, and noisy black curves in the insets) and $\alpha=1.2$ (open circles, and noisy gray curves in the insets), under stochastic variation in population sizes, $\varsigma_N=0.1\mu_N$ with $\mu_N=10^8$.
(a)~Recruitment growth-rates distributions (rescaled with $\sqrt{2}\varsigma_N/\mu_N$).
(b)~Recruitment distributions with $\expval{X}=6$ (points are binned logarithmically).
In the insets, shown are the complementary cumulative distributions.
 }\label{fig:gaussian-environment}
\end{figure}

In the following, I examine how large one can expect aggregate effects of idiosyncratic variations to be.
For $\alpha'=2$, the $N_t$'s follow the Gaussian distribution centered on $\mu_N$.
In this case, shown in Figure~\ref{fig:gaussian-environment}, one readily sees that
recruitment fluctuations in tail part come from idiosyncratic reproductive successes.
Using two different values of $\alpha=1.42$ (solid circles, and noisy black curves in the insets) and $\alpha=1.2$ (open circles, and noisy gray curves in the insets),
under stochastic variation in population sizes, $\varsigma_N=0.1\mu_N$ with $\mu_N=10^8$,
simulations were run for $10^5$ generations for each choice of parameter values, where $\expval{X}=6$.
The crossover from Gaussian to power-law behavior with exponent $\alpha$ occurs at around $\eta=\sqrt{2}{\,}\varsigma_N/\mu_N$ for recruitment growth-rate distributions (panel~a),
and the crossover also occurs at around $R/(\expval{X}\mu_N)=1+\varsigma_N/\mu_N$ for recruitment distributions (panel~b).
The black solid line in panel~a (resp. the gray dotted line in panel~b) represents the Gaussian distribution with mean zero and standard deviation $\sqrt{2}$ (resp. mean one and standard deviation $\sqrt{2}{\,}\mu_N/\varsigma_N$).
In the insets of Figure~\ref{fig:gaussian-environment}, where shown are the complementary cumulative distributions,
power-laws are seen as straight dashed lines with slopes of $-1.42$ (black) and $-1.2$ (gray) on a log-log plot.

\section{Conclusion}\label{sect:conclusion}
Based on the power-law model of offspring-number distribution with exponent $1<\alpha<2$,
this paper has provided advances in understanding the empirical properties of recruitment variability in marine fishes.
With only minimal ingredients, the L{\'e}vy-stable recruitment model is able to capture the aggregate effect of idiosyncratic variations for a large population in a random environment.
It establishes a connection between the tail exponents of the distributions at a micro and macro scale, i.e. the offspring-number and recruitment distributions.
The value of $\alpha$ may be estimated by examining the DNA sequence variation within a population, and the exponent $\alpha$ characterizes the intermittency properties (rare events of large amplitude) of recruitment
(see \citealt{Niwa-etal2016,Niwa-etal2017,Niwa-arxiv14feb2022,Niwa-arxiv10Feb2022}).

To what extent individual reproductive outputs are averaged out upon aggregation depends on the magnitude of the year-to-year fluctuations in the number of spawners or the carrying capacity of environment.
Even though the fluctuations in population sizes dominate the fluctuations of the sums of individual reproductive outputs,
the idiosyncratic reproductive successes can have a noticeable impact on the overall recruitment.
One sees the sum-stable (or $\alpha$-stable) decay
in the tails $R_t-\expval{R}\gtrsim\Sigma_R$ and $\eta_t\gtrsim\sqrt{2}{\,}\Sigma_R/\expval{R}$ with $\Sigma_R\propto\expval{R}$,
as a representation of the idiosyncratic reproductive variability,
where the tails follow the power law with the same exponent as the distribution of individual reproductive outputs.
When environmental disturbances (such as climate and commercial exploitation) are large enough for the population-size fluctuations to dominate the individual reproductive variations,
the symmetry of the L{\'e}vy-stable distribution of recruitment growth-rates is broken.
One then observes a totally asymmetric distribution with a power-law tail in a large population.

In summary, the linear-scaling behavior $\Sigma_R\propto\expval{R}$ (or the independence of $\Sigma_{\eta}$ and $\expval{R}$) implies that
the individual reproductive variability is dominated by environmentally driven fluctuations in the population size,
which leads to asymmetries in the distribution of relative changes in recruitment.
The totally asymmetric (skewed to the right) character of the recruitment growth-rate distribution can be a sign of the domination by a single or a few families in the population.
The recruitment fluctuations in tail part (i.e. intermittent, strong year-classes) come from idiosyncratic individual-level reproductive successes.

\bibliographystyle{apalike2}
\bibliography{bib-niwa-genetics}

\newcommand{\SortNoop}[1]{}
\begin{thebibliography}{}
{\footnotesize

\bibitem[Allen et~al., 2001]{Allen-etal01}
Allen, A.~P., Li, B.~L., \& Charnov, E.~L. (2001).
\newblock Population fluctuations, power laws and mixtures of lognormal
  distributions.
\newblock {\em Ecol. Lett.}, 4, 1--3.

\bibitem[Anderson et~al., 1982]{Anderson82}
Anderson, R.~M., Gordon, D.~M., Crawley, M.~J., \& Hassell, M.~P. (1982).
\newblock Variability in the abundance of animal and plant species.
\newblock {\em Nature}, 296, 245--248.

\bibitem[{\'A}rnason \& Halld{\'o}rsd{\'o}ttir,
  2015]{Arnason-Halldorsdottir2015}
{\'A}rnason, E. \& Halld{\'o}rsd{\'o}ttir, K. (2015).
\newblock Nucleotide variation and balancing selection at the \textit{{Ckma}}
  gene in {Atlantic} cod: analysis with multiple merger coalescent models.
\newblock {\em PeerJ}, 3, e786.

\bibitem[Bouchaud \& Georges, 1990]{Bouchaud-Georges90}
Bouchaud, J.~P. \& Georges, A. (1990).
\newblock Anomalous diffusion in disordered media: statistical mechanisms,
  models and physical applications.
\newblock {\em Phys. Rep.}, 195, 127--293.

\bibitem[Carmona \& Durrleman, 2003]{Carmona-Durrleman2003}
Carmona, R. \& Durrleman, V. (2003).
\newblock Pricing and hedging spread options.
\newblock {\em SIAM Rev.}, 45, 627--685.

\bibitem[Cobain et~al., 2019]{Cobain2019}
Cobain, M. R.~D., Brede, M., \& Trueman, C.~N. (2019).
\newblock Taylor's power law captures the effects of environmental variability
  on community structure: an example from fishes in the {North Sea}.
\newblock {\em J. Anim. Ecol.}, 88, 290--301.

\bibitem[Cushing, 1990]{Cushing1990}
Cushing, D. (1990).
\newblock Plankton production and year-class strength in fish populations: an
  update of the match/mismatch hypothesis.
\newblock {\em Adv. Mar. Biol.}, 26, 249--293.

\bibitem[Eisler et~al., 2008]{Eisler2008}
Eisler, Z., Bartos, I., \& Kert{\'e}sz, J. (2008).
\newblock Fluctuation scaling in complex systems: {Taylor's} law and beyond.
\newblock {\em Adv. Phys.}, 57, 89--142.

\bibitem[Eldon et~al., 2015]{EBBF2015}
Eldon, B., Birkner, M., Blath, J., \& Freund, F. (2015).
\newblock Can the site-frequency spectrum distinguish exponential population
  growth from multiple-merger coalescents?
\newblock {\em Genetics}, 199, 841--856.

\bibitem[Eldon \& Wakeley, 2006]{EW06}
Eldon, B. \& Wakeley, J. (2006).
\newblock Coalescent processes when the distribution of offspring number among
  individuals is highly skewed.
\newblock {\em Genetics}, 172, 2621--2633.

\bibitem[Gabaix, 2011]{Gabaix2011}
Gabaix, X. (2011).
\newblock The granular origins of aggregate fluctuations.
\newblock {\em Econometrica}, 79, 733--772.

\bibitem[Gibrat, 1931]{Gibrat31}
Gibrat, R. (1931).
\newblock {\em Les In{\'e}galit{\'e}s {\'E}conomiques}.
\newblock Paris: Libraire du Recueil Sirey.

\bibitem[Hedgecock \& Pudovkin, 2011]{Hedgecock-Pudovkin2011}
Hedgecock, D. \& Pudovkin, A.~I. (2011).
\newblock Sweepstakes reproductive success in highly fecund marine fish and
  shellfish: a review and commentary.
\newblock {\em Bull. Mar. Sci.}, 87, 971--1002.

\bibitem[Hilborn \& Walters, 1992]{Hilborn-Walters92}
Hilborn, R. \& Walters, C.~J. (1992).
\newblock {\em Quantitative Fisheries Stock Assessment: Choice, Dynamics and
  Uncertainty}.
\newblock London: Chapman \& Hall.

\bibitem[Hjort, 1914]{Hjort1914}
Hjort, J. (1914).
\newblock Fluctuations in the great fisheries of the northern {Europe} viewed
  in the light of biological research.
\newblock {\em Rapp. P.-V. R{\'e}un. Cons. Int. Explor. Mer}, 20, 1--228.

\bibitem[Keitt \& Stanley, 1998]{Keitt-Stanley98}
Keitt, T. \& Stanley, H.~E. (1998).
\newblock Dynamics of {N}orth {A}merican breeding bird populations.
\newblock {\em Nature}, 393, 257--260.

\bibitem[Keitt et~al., 2002]{Keitt-etal02}
Keitt, T.~H., Amaral, L. A.~N., Buldyrev, S.~V., \& Stanley, H.~E. (2002).
\newblock Scaling in the growth of geographically subdivided populations:
  invariant patterns from a continent-wide biological survey.
\newblock {\em Phil. Trans. R. Soc. Lond. B}, 357, 627--633.

\bibitem[Khintchine \& L{\'e}vy, 1936]{Khintchine-Levy1936}
Khintchine, A. \& L{\'e}vy, P. (1936).
\newblock Sur les lois stables.
\newblock {\em C. R. Acad. Sci. Paris}, 202, 374--376.

\bibitem[Kingman, 1982]{Kingman82-the-coalescent}
Kingman, J. F.~C. (1982).
\newblock The coalescent.
\newblock {\em Stoch. Process. Their Appl.}, 13, 235--248.

\bibitem[Lan \& Chandran, 2011]{Lan2011}
Lan, B.~L. \& Chandran, P. (2011).
\newblock Distribution of animal population fluctuations.
\newblock {\em Physica A}, 390, 1289--1294.

\bibitem[L{\'e}vy, 1925]{Levy1925}
L{\'e}vy, P. (1925).
\newblock {\em Calcul des Probabilit{\'e}s}.
\newblock Paris: Gauthier-Villars.

\bibitem[L{\'e}vy, 1937]{Levy1937}
L{\'e}vy, P. (1937).
\newblock {\em Th{\'e}eorie de l'Addition des Variables Al{\'e}atoires}.
\newblock Paris: Gauthier-Villars.

\bibitem[{\SortNoop{Menezes}}de~Menezes \& Barab{\'a}si,
  2004]{deMenezes-Barabasi92}
{\SortNoop{Menezes}}de~Menezes, M.~A. \& Barab{\'a}si, A.-L. (2004).
\newblock Fluctuations in network dynamics.
\newblock {\em Phys. Rev. Lett.}, 92, 028701.

\bibitem[Newman, 2005]{Newman2005}
Newman, M. E.~J. (2005).
\newblock Power laws, {Pareto} distributions and {Zipf's} law.
\newblock {\em Contemp. Phys.}, 46, 323--351.

\bibitem[Niwa, 2006]{Niwa06-ecoinf}
Niwa, H.-S. (2006).
\newblock Exploitation dynamics of fish stocks.
\newblock {\em Ecol. Inform.}, 1, 87--99.

\bibitem[Niwa, 2007]{Niwa07-ICES}
Niwa, H.-S. (2007).
\newblock Random-walk dynamics of exploited fish populations.
\newblock {\em ICES J. Mar. Sci.}, 64, 496--502.

\bibitem[Niwa, 2022a]{Niwa-arxiv11feb2022}
Niwa, H.-S. (2022a).
\newblock Fluctuation scaling in {L{\'e}vy-stable} recruitment of marine fishes
  in randomly varying environments.
\newblock {\em arXiv:2202.06206 [q-bio.PE]}.

\bibitem[Niwa, 2022b]{Niwa-arxiv14feb2022}
Niwa, H.-S. (2022b).
\newblock Hatchery-induced transition of the effective size in a {Pareto}
  population.
\newblock {\em arXiv:2202.07038 [q-bio.PE]}.

\bibitem[Niwa, 2022c]{Niwa-arxiv10Feb2022}
Niwa, H.-S. (2022c).
\newblock Reciprocal symmetry breaking in {Pareto} sampling.
\newblock {\em arXiv:2202.04865 [math.PR]}.

\bibitem[Niwa et~al., 2016]{Niwa-etal2016}
Niwa, H.-S., Nashida, K., \& Yanagimoto, T. (2016).
\newblock Reproductive skew in {Japanese} sardine inferred from {DNA}
  sequences.
\newblock {\em ICES J. Mar. Sci.}, 73, 2181--2189.

\bibitem[Niwa et~al., 2017]{Niwa-etal2017}
Niwa, H.-S., Nashida, K., \& Yanagimoto, T. (2017).
\newblock Allelic inflation in depleted fish populations with low recruitment.
\newblock {\em ICES J. Mar. Sci.}, 74, 1639--1647.

\bibitem[Reed \& Hughes, 2002]{Reed-Hughes2002}
Reed, W.~J. \& Hughes, B.~D. (2002).
\newblock From gene families and genera to incomes and internet file sizes: why
  power laws are so common in nature.
\newblock {\em Phys. Rev. E}, 66, 067103.

\bibitem[Sargsyan \& Wakeley, 2008]{Sargsyan-Wakeley08}
Sargsyan, O. \& Wakeley, J. (2008).
\newblock A coalescent process with simultaneous multiple mergers for
  approximating the gene genealogies of many marine organisms.
\newblock {\em Theor. Popul. Biol.}, 74, 105--114.

\bibitem[Schweinsberg, 2003]{Schweinsberg2003}
Schweinsberg, J. (2003).
\newblock Coalescent processes obtained from supercritical {Galton-Watson}
  processes.
\newblock {\em Stoch. Process. Their Appl.}, 106, 107--139.

\bibitem[Steinr{\"u}cken et~al., 2013]{SBB13}
Steinr{\"u}cken, M., Birkner, M., \& Blath, J. (2013).
\newblock Analysis of {DNA} sequence variation within marine species using
  {Beta}-coalescents.
\newblock {\em Theor. Popul. Biol.}, 87, 15--24.

\bibitem[Takayasu et~al., 2014]{Takayasu-etal2014}
Takayasu, M., Watanabe, H., \& Takayasu, H. (2014).
\newblock Generalised central limit theorems for growth rate distribution of
  complex systems.
\newblock {\em J. Stat. Phys.}, 155, 47--71.

\bibitem[Taylor, 1961]{Taylor61}
Taylor, L.~R. (1961).
\newblock Aggregation, variance and the mean.
\newblock {\em Nature}, 189, 732--735.

\bibitem[Zaliapin et~al., 2005]{ZKS05}
Zaliapin, I.~V., Kagan, Y.~Y., \& Schoenberg, F.~P. (2005).
\newblock Approximating the distribution of {Pareto} sums.
\newblock {\em Pure Appl. Geophys.}, 162, 1187--1228.

}
\end{thebibliography}

\end{document}